# Graphene/g-C$_2$N bilayer: gap opening, enhanced visible light response and electrical field tuning band structure

*Zhaoyong Guan[a], Jia Li[b] and Wenhui Duan[acd]*

[a]Department of Physics, Tsinghua University, Beijing 100084, P. R. China

[b]Guangdong Key Laboratory of Thermal Management Engineering and Materials, Graduate School at Shenzhen, Tsinghua University, Shenzhen 518055, P. R. China

[c]Collaborative Innovation Center of Quantum Matter, Tsinghua University, Beijing 100084, P. R. China

[d]Institute for Advanced Study, Tsinghua University, Beijing 100084, P. R. China

**ABSTRCT**

Opening up a band gap of the graphene and finding a suitable substrate are two challenges for constituting the nano-electronic equipment. A new two-dimensional layered crystal g-C$_2$N (Nat. Commun. 2015, 6, 1–7) with novel electronic and optical properties can be effectively synthesized via a wet-chemical reaction. And g-C$_2$N can be used as a suitable substrate to open the band gap of graphene as much as 0.239 eV, which is large enough for the band gap opening at room temperature. The physics behind the band gap opening is that g-C$_2$N substrate can produce the inhomogeneous electrostatic potential over the graphene layer. The imposition of external electrical field can tune the band gap of the hybrid of graphene/g-C$_2$N effectively from the semiconductors to the metal. The hybrid graphene/g-C$_2$N displays an enhanced optical activity compared with the pure g-C$_2$N monolayer.

**Introduction**

Graphene is composed of a single layer of carbon atoms arranged in a two-dimensional (2D) honeycomb lattice,[1] which attracts much attention for its special properties.[2-5] Many extraordinary properties of graphene, such as 2.3% absorption in the white light spectrum, high surface area, high Young's modulus, and excellent thermal conductivity, have been extensively investigated.[6] One of the most interesting aspects of the graphene's property is its low-energy excitations being massless, chair, Dirac fermions. And Dirac fermions behave in unusual ways as compared to ordinary electrons when subjected to magnetic fields, leading to new physical phenomena, such as the anomalous integer quantum Hall effect.[7-10] Unfortunately, the lack of a finite band gap implies that the current can never be tuned off completely, which is formidable hurdle to the use of graphene in logic and high speed switching devices.[11-12] So researchers attempted to find methods to open the band gap of the graphene for its application.[13-18] These methods include cutting the graphene into all kinds of chirality of the nano-ribbons and single-wall nanotubes,[13,14] stretching the graphene with the uniaxial strain,[15,16] chemically functioning with O, H and halogen atoms,[17,18] epitaxial growing on the all kinds of the substrates,[19] and doping with molecules.[20] Besides that, applying an external electrical field to a bilayer graphene could also open a band gap,[21,22] but it is hard to be realized and effectively controlled in the experiment.[11,23] For the graphene based electronic devices, the electronic properties and high electron mobility could be tuned by the substrate. It is important to find the proper substrate. We now face another challenge of finding the ideal substrates on which graphene can keep superior properties.[24] The most commonly used substrates are SiC,[25,26] SiO$_2$[27,28] and other metals such as Ni, but they are never the suitable substrate for their high surface roughness and optical phonons.[29] Comparing with other substrates, hexagonal boron nitride shows little lattice mismatch and has rather smooth surface. But the effects depend strongly on their stacking order, which cannot be easily controlled in the experiments.[30,31]

Very recently, a thin layered g-$C_2$N (designated as $C_2$N-h2D crystal) has been successfully synthesized with a bottom-up wet-chemical reaction, and confirmed experimentally with various characterization methods.[32] The $C_2$N-h2d crystal has distributed holes and nitrogen atoms in the layered structure with the symmetry group of D6H-1. It shows $sp^2$ hybridization characters, as semiconductor with 1.96 eV band gap. In the FET devices, it shows a $10^7$ on/off ration, which is much higher than the graphene. These characters make the g-$C_2$N potentially a very useful material for future application in optoelectronic conversion and as a photocatalyst for water splitting into hydrogen.[32] In the early work, although the stacking of g-$C_3N_4$ and Graphene has been synthesized in the experiment, the g-$C_3N_4$ open the gap of the graphene only as small as 0.07 eV with quite low on/off ration.[33,34] The focuses are now on the following questions: (i) Is there any substrate such as g-$C_2$N can open bigger gap of graphene for the high on/off ration? (ii) Can the optical properties be effectively increased? and (iii) How the electrical field tune the electric properties of the graphene/g-$C_2$N complex?

In this article, using first-principles method and long-range dispersion correction[35], we calculate structural and electronic properties of the hybrid graphene/g-$C_2$N nano-composite. We prove that the g-$C_2$N can open the band gap of the graphene for the first time. Furthermore, the results of optical adsorption spectrum show obvious enhanced visible light response. The external electrical field (E-field) can efficiently tune the electronic properties of hybrid graphene/g-$C_2$N complex.

**Computational methods**

The calculations on the hybrid graphene/g-C$_3$N$_4$ complex are performed by using the plane-wave basis Vienna Ab initio Simulation Package (VASP) code,[37,38] based on density functional theory under the generalized gradient approximation (GGA) with the Perdew-Burke-Ernzerhof (PBE).[36] In order to accurately describe the weak van der Waals interactions, a damped van der Waals (vdW) correction (PBE-D2)[39] developed by Grimme has been adopted, which could provide a good description of long-range vdW interactions.[40-42] The calculated interlayer distance for bilayer graphene is 3.25 Å. The binding energy per carbon atom is about -25 meV, which is consistent with the previous experiment measurement and theoretical studies.[43-45] The vacuum space in the z-direction is set as large as 15 Å to avoid interactions between periodic images. The cutoff kinetic energies for plane waves is set as 520 eV, and the geometry structures are fully relaxed until energy and forces are converged to $10^{-5}$ eV and 0.01 eV/Å. Brillouin zone integration is used adopted 1×1×1, 1×1×1, and 6×6×1 Monkhorst-Pack k-grids for optimizing structure, computing energy and density of the states. And 90 uniform k-points along the one-dimensional Brillouin zone are used to obtain the band structure. Optimizing geometry without restricting symmetry are performed by using the conjugate gradient scheme until the force acting on every atoms is less than 10 meV/Å.

**Results and discussion**

In the calculation, we use supercell model to describe the graphene/g-C$_2$N hetero-structure. For the little lattice mismatch, we use quite large supercell, which includes 7×7×1 graphene supercell and 2×2×1 g-C$_2$N supercell to match with each other. The total supercell contains 170 atoms, including 146 C atoms and 24 N atoms. The lattice parameter of g-C$_2$N and graphene is 8.323 Å and 2.460 Å, respectively. For the adopted bigger supercell, there's only 3.3% lattice mismatch. The use of such big supercell decreases the strain effect. Besides, the artificial strain introduced to match the lattice parameters does not affect the main conclusions. In the calculation, we enlarge the lattice of the g-$C_2N$ with the 3.3% biaxial strain. And we also test the effect of the enlarged and compressed strain on the electronic property. The band gap only change slightly when change the lattice parameters of the g-$C_2N$. Therefore, the lattice parameter is also enlarged by 3.3% to match the lattice of the graphene. First, we recalculate the band gap of the g-$C_2N$ as the benchmark. The g-$C_2N$ is direct semiconductor with 1.66 eV gap at the $\Gamma$ point, which is consistent with the previous results.[32] These results illustrate that the parameters are reasonable and reliable. We also calculate the band structure of the graphene, and there are still K and K` located at the (0.33333, 0.66667, 0.00000) and (-0.33333, 0.66667, 0.00000), respectively. The C-C bond length of the optimized structure is about 1.42 Å, which is consistent with the experimental and theoretical results.[2]

The effect of strain on the electronic properties is also taken into consideration. Neither compressed strain nor enlarged strain has small effect on the band structure of the g-$C_2N$, with the corresponding results shown in the supporting information. The optimized geometry and band structure of the g-$C_2N$ are shown in Fig. 1. The corresponding bond lengths of the C-N and C-C are 1.336 Å and 1.430 Å, respectively. And g-$C_2N$ is a layered two-dimensional network structure possessed evenly distributed holes. All the atoms are nearly in the same plane, which is consistent with the previous results. [32]

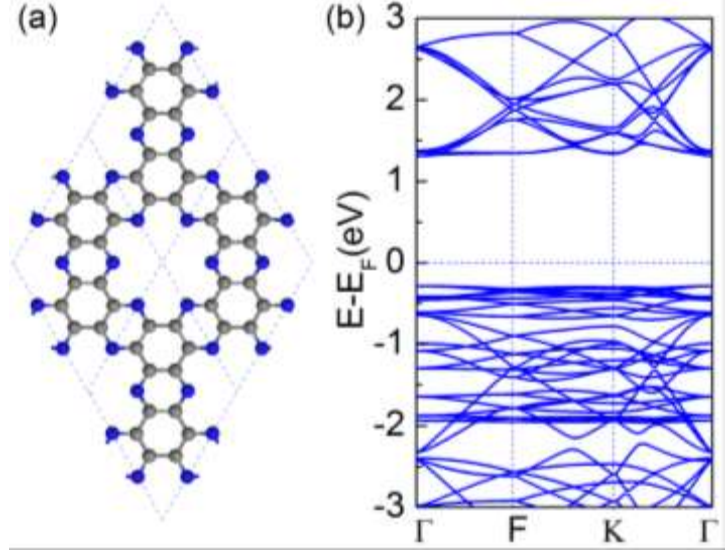

Figure 1. (a) Optimized structure of a g-$C_2N$ monolayer, (b) band structure of g-$C_2N$ monolayer calculated by PBE. The grey and blue balls represent C and N atoms, respectively. The $\Gamma$, F, K presents (0.00000, 0.00000, 0.00000), (0.00000, 0.50000, 0.00000), K (0.33333, 0.66667, 0.00000), respectively.

We stack graphene and g-$C_2N$ in the z-direction and fully relax geometry of hybrid graphene/g-$C_2N$. The calculated equilibrium distance between the graphene layer and the g-$C_2N$ monolayer is 3.26 Å, which is closer the distance between the bilayer of the g-$C_2N$ (3.28 Å).[32] The absorbed energy was obtained according to the following equation:

$$E_{ab} = E_{graphene/g-C_2N} - E_{graphene} - E_{g-C_2N}$$

where $E_{graphene/g-C_2N}$, $E_{graphene}$ and $E_{g-C_2N}$ stands for the total energy of the hybrid graphene and g-$C_2N$ complex, pure graphene and pure g-$C_2N$, respectively. The adsorbed energy is as high as -0.85 eV for the entire system, which indicates very highly stable vdW interactions among them as observed in the experiments.[32] This energy is calculated with the PBE with long-range dispersion correction. The vdW interaction plays an important role in describing the non-bond interaction.

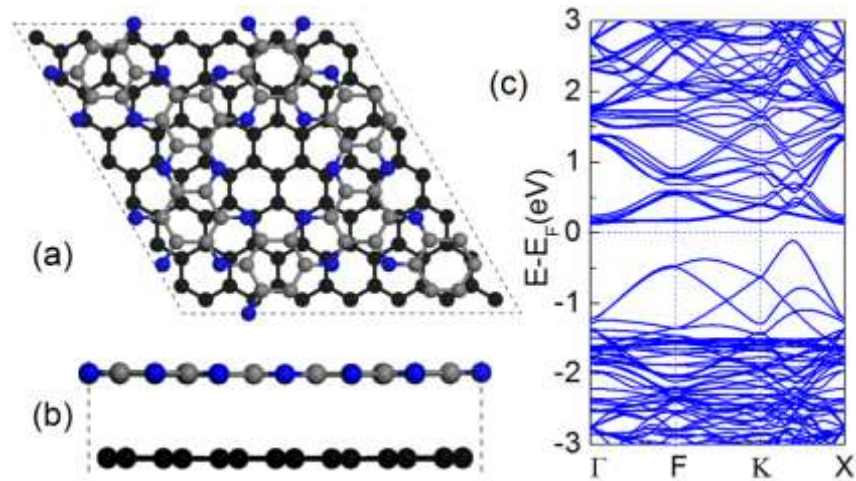

Figure 2. Optimized hybrid graphene/g-$C_2N$ composites in a (a) top view and (b) side view, (c) band structure calculated by the PBE functional for the hybrid graphene/g-$C_2N$ complex. Black, grey and blue balls represent C atoms of the graphene, C and N atoms of the g-C2N monolayer, respectively.

The optimized structure and band structure of the hybrid graphene/g-$C_2N$ complex are shown in Fig. 2. For graphene/g-$C_2N$ composites, the linear band structure is disrupted, as compared with that of the pristine graphene, to form metal-to-semiconductor translation. Although the Dirac cone remains, it lays above the fermi-level instead of crossing. However, there is gap of 0.780 eV opened at the K point of the Brillouin Zone. The hybrid graphene/g-$C_2N$ composite behaves as an indirect semiconductor with gap of 0.239 eV. The valence band maximum (VMB) and the conduction band minimum (CBM) are located respectively at the M (0.21900, 0.43700, 0.00000) and Γ (0.00000, 0.00000, 0.00000) points of the Brillouin Zone. For the original band structure of the g-$C_2N$, the VBM and CBM both lay at the Γ point. In order to investigate the origins for opening the band gap of the graphene, we examine the electrical potential of the g-$C_2N$ (see Figure S3 in the supplements). The g-$C_2N$ creates an inhomogeneous electrical potential across the space above the graphene. This changes the periodic potential and degenerates the π and π` bands at the K (0.33333, 0.66666, 0.00000) point. Besides, the graphene and the g-$C_2$N maintain the perfect planer structures.

The graphene is semimetal with a band gap of 0 eV, which inhibits the graphene from being used in the FET. [46] The present hybrid graphene/g-$C_2$N composites can be a candidate material for the FET. For the hybrid graphene/g-$C_2$N composites, the current on-off ratio can be effectively controlled by tuning the voltage of a back-gate.[11] Most importantly the band gap the hybrid graphene/g-$C_2$N composites (0.239 eV) is far larger than the room-temperature thermal-energy ($K_B$T), that insures the high on-off ratio. We have calculated the band structure, with the total and partial density of the states (DOS) shown in Fig. 3. These results imply that both the g-$C_2$N and graphene states contribute to the states near the Fermi level. Therefore, the states near the Fermi level consist mainly of 2$p_z$ orbitals of the graphene and g-$C_2$N (see the Figure S4 in the supporting information), which is the same with the hybrid graphene/g-$C_3N_4$ composites. [11]

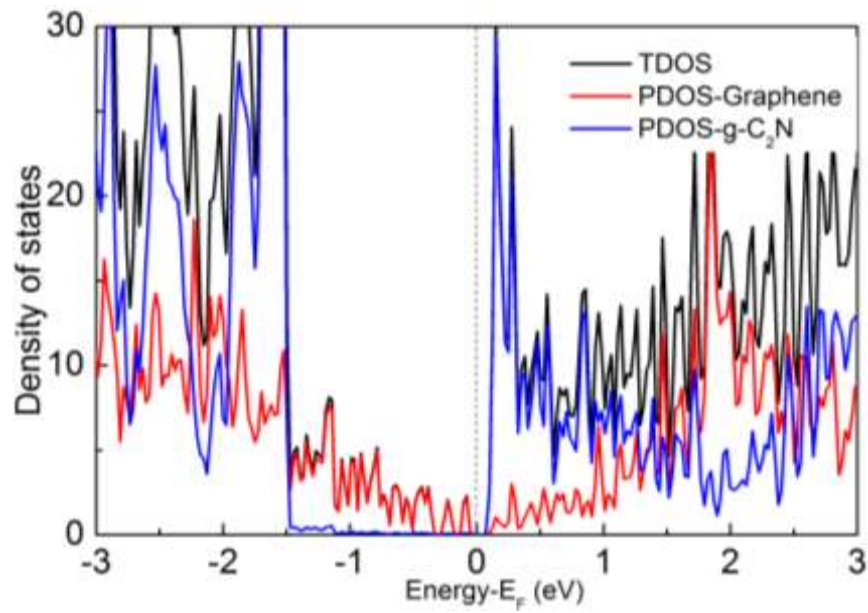

Figure 3. The density of the states of the hybrid graphene/ g-$C_2$N complex. The blue line represents the total DOS, while black and red lines the PDOS of g-$C_2$N and graphene, respectively.

To provide the reason for the opening the band gap and the coupling between the graphene and g-$C_2N$, we plot the charge density difference between the graphene and the g-$C_2N$. The side and top views of the difference of charge density are shown in Figs. 4 (a) and (b). From these figures, the charge density is redistributed by forming electron-rich and hole-rich regions with the graphene. The similar phenomena also appear in the hybrid of graphene/g-$C_3N_4$ composites.[11, 34]

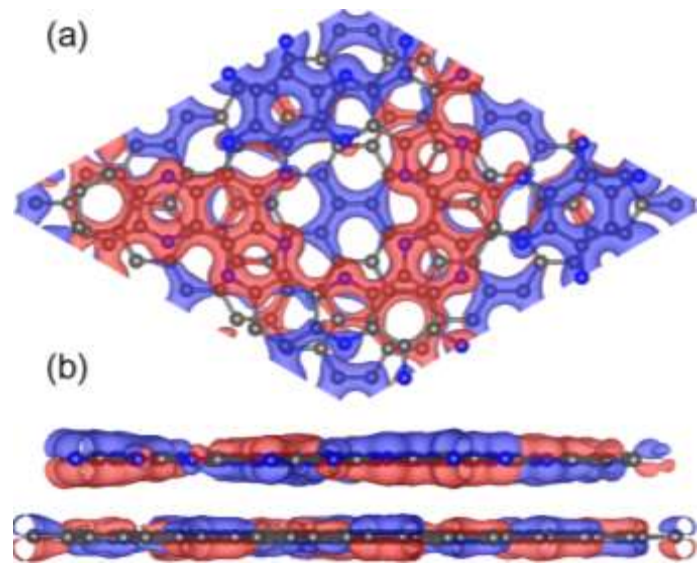

Figure 4. (a) Top view and (b) side view of the charge density difference of the hybrid of graphene/g-$C_2N$ complex. The grey and black balls represent C and N atoms, respectively. The isovalue chosen to plot the isosurface is 0.002 e Å$^{-3}$.

The origin of opening the band gap of the graphene is the different chemical environment (the different electrostatic potential over the whole graphene layer) of the two sub-lattices, which in term is originated from the charge difference of hybrid of graphene/g-$C_2N$. The vertical electrical field (E-field) can effectively modulate the charge density difference. And the external electric field could affect the bandgap of 2D materials. As a result, we consider the use of E-field to tune the electronic structure of the hybrid graphene/g-$C_2N$.

The effect of E-field on the binding energy and electronic properties of the hybrid graphene/g-$C_2N$ composites has been carefully investigated. The E-filed is in the z-direction perpendicular to the planes of graphene and g-$C_2N$, whose strength changes from the -0.6 to 0.6 V/Å. The positive direction of the E-field points from the graphene to the g-$C_2N$. The E-field can tune the structural properties, bonding strength and electronic properties of the hybrid graphene/g-$C_2N$, as well as the distance between the graphene and g-$C_2N$. Without the E-field (E=0 V/Å), the distance between the graphene and g-$C_2N$ is 3.251 Å, which is used as reference. As the E-field increases along the positive z direction, the distance decreases. For weaker electrical fields of E=0.1, 0.2, 0.3 and 0.4 V/Å, the distances are 3.251, 3.250, 3.250 and 3.249 Å, respectively. For the stronger electrical field of E=0.5, and 0.6 V/Å, the distances are 3.238 and 3.234 Å, respectively. When the direction of the E-field is reversed, the distance decreases as the strength of the E-field increases. For E= -0.1, -0.2, -0.3, -0.4 and -0.5, the corresponding distances are 3.250, 3.250, 3.250, 3.251 and 3.251 Å, respectively. For E=-0.6 V/Å, the distance is 3.258 Å. In a word, the E-field could tune the distance between the graphene and g-$C_2N$. The binding energy often changes with the distance between the two layers,[34] so we also calculate the binding energy under the different E-fields. We define the binding energy of the supercell as the $E_{ab}/N$, where $E_{ab}$ stands for the total binding energy of the hybrid graphene/g-$C_2N$, and N the number of the units of graphene in the supercell (Here N=49 of this study). As indicated from the binding energies shown in Fig. 5, the interaction between graphene and g-$C_2N$ could be enhanced by E-field. For E=0 V/Å, the binding energy of per cell is -0.074 eV. For E=-0.2, -0.4 and -0.6, the corresponding binding energies are -0.076, -0.081 and -0.092 eV, respectively. For the E=0.2, 0.4 and 0.6, the corresponding binding energies are -0.078, -0.087 and -0.101, respectively. From the above results, we conclude that the binding energy increases as the strength of E-field increase, irrespective to the direction of the E-field.

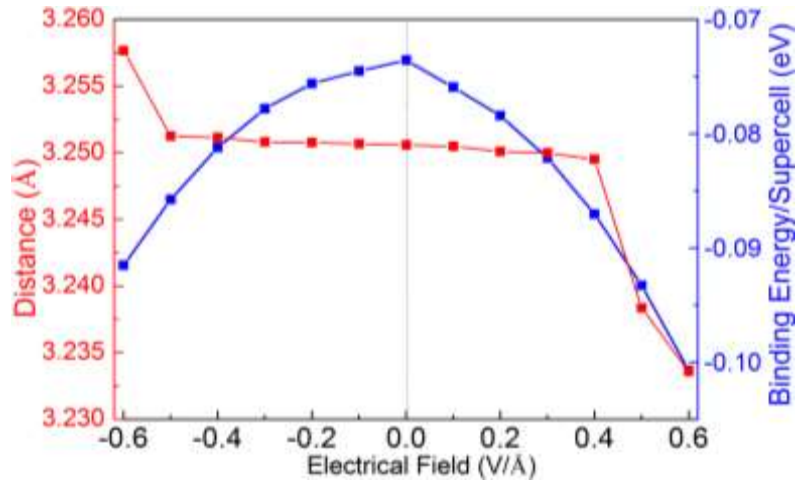

Figure 5. The distance between the graphene and g-$C_2N$ bilayer and binding energy per supercell changes with the E-field.

In the above, we have investigate the effects of E-field on structural properties and binding energy of the hybrid of the graphene/g-$C_2N$.. Below we systematically investigate the change in the electronic properties associated with the structure change. It has been demonstrated that the external electric field could affect the bandgap of 2D materials in previous theoretical and experiment work.[47-52] The change in band gaps due to the imposition of E-field are shown in Fig. 6. For the E-field along the negative z direction, the gaps increase with increasing strength of E-field, similarly for the absolute value of VBM. On the other hand, the value of CBM increases first with increasing E-field and then decrease as the strength of E-field increases further. It is interesting to note that the band gap changes linearly with the E-field in the range of (-0.2, 0.3), which can be fitted with the following relation, Egap= 0.266-0.947E, for the practical use in the nano-electronics. As the E-field increases along the positive z-direction, the gap decreases to 0 eV at E=0.3V/Å, where the system converts from semiconductor into metal.

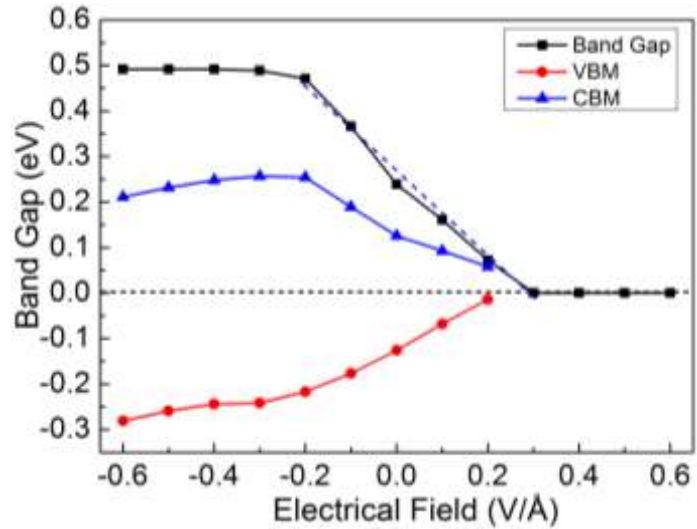

Figure 6. Band gaps of the hybrid graphene/g-C$_2$N change as function of the E-field. The black, red and blue line shows the band gap, VBM and CBM, respectively.

For the stronger E-field, the system remains as metal. The band structures with E-field= -0.6, -0.4, -0.2, 0.2, 0.4 and 0.6 V/Å are shown in Fig. 7. From Fig. 7, the morphologies of the band structure under the E-field along the negative z-direction cannot change dramatically except for the gap difference. The band gap modulation of the hybrid graphene/g-C$_2$N complex by E-field as discussed above could be explained with the Stark effect. Single-walled boron nitride nanotubes[55] and nanoribbons[56] and MoS2 bilayer[57] show the similar phenomenon. Under the external E-field, both VBM and CBM change with the strength of E-field, and finally cause a transition from semiconductor to metal.

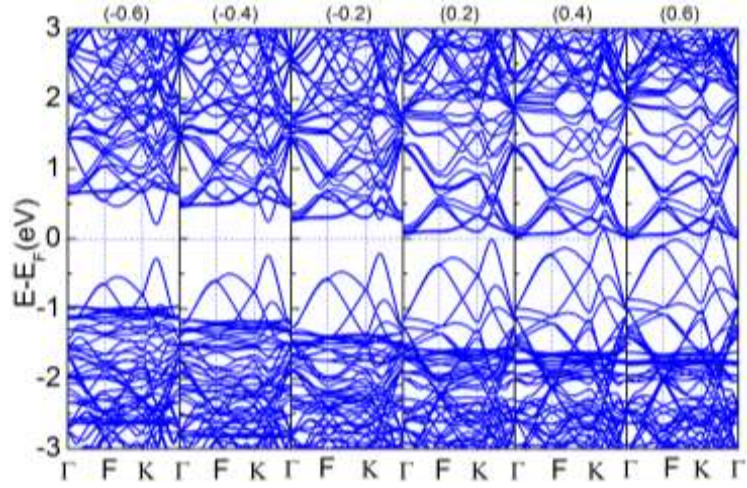

Figure 7. The band structure under different E-field= -0.6, -0.4, -0.2, 0.2, 0.4, 0.6 V/ Å, respectively.

We will discuss in the following the photo absorption efficiency of the hybrid graphene/g-$C_2N$. Firstly, the frequency-dependent dielectric function, $\varepsilon(\omega) = \varepsilon_1(\omega) + i\varepsilon_2(\omega)$, is calculated, and the absorption coefficient as a function of photon energy is evaluated with the following expression:

$$\alpha(E) = \frac{4\pi e}{hc}\left(\frac{\left[\varepsilon_1^2 + \varepsilon_2^2\right]^{1/2} - \varepsilon_1}{2}\right)^{1/2}$$

The imaginary part of the dielectric function for g-$C_2N$ and hybrid graphene/g-$C_2N$ is calculated by the PBE, and the results are shown in Figs. 8. (a) and (b). Different absorption behaviors appear in the transversal and longitudinal directions, as a result of structural anisotropy. The absorption in longitudinal direction is much stronger than that in the transversal direction, to construct the main contribution to the whole absorption. Similar phenomenon also occurs in the semi-hydrogenated BN sheet.[58] Comparing with the pure g-$C_2N$, the efficiency of photo absorption of hybrid graphene/g-$C_2N$ is obviously enhanced and efficiently improved for the narrower band gap. For the hybrid graphene/g-$C_2N$, there is an obvious blue shift. The first absorption peek shifts from the 2.433 eV of the pure g-$C_2N$ to 0.734 eV of the hybrid graphene/g-$C_2N$, with the absorption coefficient being as high as $10^5$. Although the PBE functional often underestimates the band gap and could not describe the absorption coefficient accurately, the trend is still applicable.

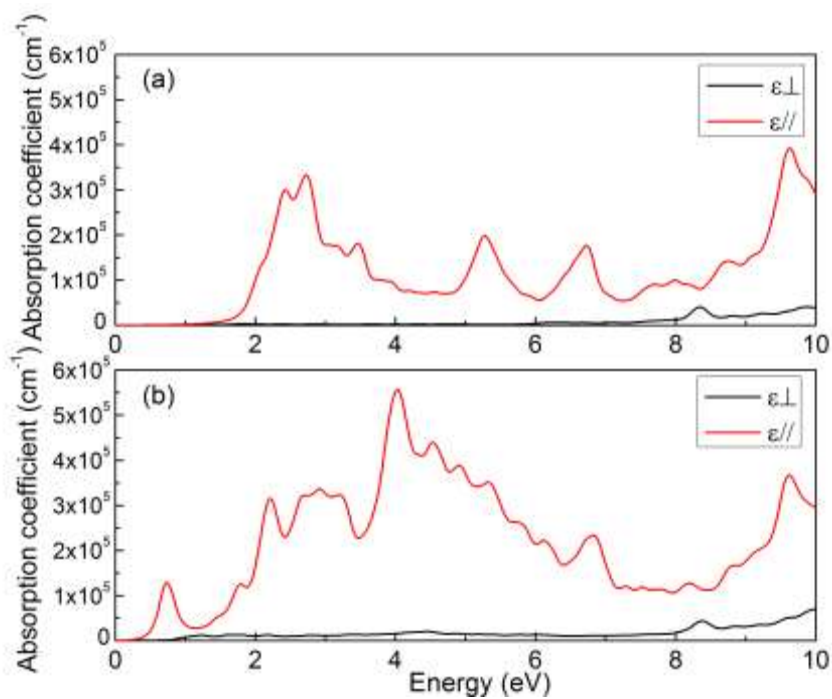

Figure 8. Optical spectra for (a) pure g-$C_2N$ monolayer and (b) hybrid graphene/g-$C_2N$. The black and red line stands for the absorption coefficient perpendicular and parallel to the z-direction, respectively.

**Conclusions**

We have for the first time investigated comprehensively the structural and electronic properties of the hybrid graphene/g-$C_2N$ with the first principle. The band gap of hybrid graphene/g-$C_2N$ is large enough (0.239 eV) to overcome the thermal excitation at room temperature, to remove the main hinder of graphene used in novel integrated functional nano-devices. Furthermore, the band gaps of the hybrid graphene/g-$C_2N$ could be effectively tuned by the E-field according to the Stark effect. The hybrid graphene/g-$C_2N$ complex also shows an enhanced optical absorption behavior, making it a candidate material as metal-free photo-catalysts and photovoltaic devices.

**AUTHOR INFORMATION**

**Corresponding Author**

* E-mail: lijia@phys.tsinghua.edu.cn. Phone: +86-755-26033022. Fax: +86-755-26036417 (J. L.). dwh@phys.tsinghua.edu.cn. Phone: +86-010-62785577. Fax: +86- 010-62781604 (W. H.).

**Author Contributions**

The manuscript was written through contributions of all authors. All authors have given approval to the final version of the manuscript.

**ACKNOWLEDGMENT**

We thank Shuanglin Hu and Ruiqi Zhang for useful discussions. This work is partially supported by National Key Basic Research Program, and by USTCSCC, SCCAS, Tianjin, and Shanghai Supercomputer Centers.